\documentclass{aa}

\begin{document}

   \thesaurus{ 06      
              (02.13.3;  
               02.16.2;  
               03.20.2;  
               08.16.4;  
               13.19.5)}  
   \title{VLBA observations of SiO masers: Arguments in favor of
	  radiative pumping mechanisms }


   \author{	J.F. Desmurs\inst{1,2}
	\and	V. Bujarrabal\inst{1}
	\and	F. Colomer\inst{1}
	\and	J. Alcolea\inst{1}
	}

   \offprints{J.F. Desmurs}

   \institute{Observatorio Astron\'omico Nacional (IGN), Apartado 1143,
		E--28800 Alcal\'a de Henares, Spain \\
              	email: desmurs,bujarrabal,colomer,j.alcolea@oan.es 
         \and
		Joint Institute for VLBI in Europe, Postbus 2, 
		NL-7990 AA Dwingeloo, The Netherlands \\
             }

   \date{H1543, Received 8/01/99; Accepted 09/06/2000}
   \titlerunning {VLBA observations of SiO masers}
   \maketitle

   \begin{abstract}

We have performed VLBA observations of the SiO $v$=1 and $v$=2 $J$=1--0
masers in two AGB stars, TX~Cam and IRC~+10011. We confirm the
ring-like spatial distribution, previously found in several AGB
objects, as well as the tangential polarization pattern, already
reported for TX~Cam.  Both properties, that seem to be systematic in
this kind of objects, are characteristic of radiatively pumped SiO
masers. On the contrary, we do not confirm the previous report on the
spatial coincidence between the $J$=1--0 $v$=1 and 2 masers, a result
that would have argued in favor of collisional pumping. We find that
both lines sometimes arise from nearby spots, typically separated by
1--2 mas, but are rarely coincident. The discrepancy with previous
results is explained by the very high spatial resolution of our
observations, $\sim$ 0.5 mas, an order of magnitude better than in the
relevant previously published experiment.


   \end{abstract}


\section{Introduction}
\label{intro}

SiO maser emission at 7 mm wavelength ($v$=1 and $v$=2, $J$=1--0
transitions near 43~GHz) has been observed in AGB stars with very high
spatial resolution by means of VLBI techniques, yielding important
results in relation with their not yet well understood pumping
mechanism.  The 7~mm maser emission regions are found to be distributed
in a number of spots forming a ring-like structure at about 2--3
stellar radii: see observations of TX~Cam, U\,Her, W\,Hya, and VX\,Sgr
by Diamond et al$.$\ (1994), Miyoshi et al$.$\ (1994), and Greenhill et
al$.$\ (1995).  This ring-like flux distribution arises naturally in
the framework of the radiative pumping mechanism of SiO masers (see
\cite{buja94a}, and references therein), that requires a shell-like
matter distribution.  These structures may also be possible in
collisional models, which are practically not geometry dependent
(e.g$.$ \cite{doel95}).  From observations with the KNIFE VLBI array,
Miyoshi et al$.$\ (1994) reported that the emission of these two lines
systematically arises from positionally coincident spots, at least as
seen with their 7~mas beam. Miyoshi et al$.$\ claimed that this result
is a conclusive proof in favor of collisional pumping schemes, since
radiative mechanisms, always more selective, tend to require different
physical conditions to pump the $v$=1 and $v$=2 masers.

VLBA observations (\cite{kemb97}) have also shown that most SiO spots
are linearly polarized in the tangential direction, i.e$.$
perpendicularly to the direction to the ring center, in a particularly
well defined pattern. This result was interpreted as due to a magnetic
field, but uncomfortably strong (close to 10~G). Moreover, Kemball \&
Diamond found problems to define a distribution of possible field
directions that would explain such a tangential pattern, mostly if it
systematically applies to SiO masers in AGB stars.
The problem of explaining the tangential polarization can be solved
invoking radiative pumping. Tangential linear polarization in SiO
masers arises in these models as a direct result of the anisotropy
introduced by the absorption of stellar (radial) photons, even in the
absence of magnetic field. In fact, tangential polarization was
predicted from radiative pumping models more than ten years before any
VLBA observations were available (Bujarrabal \& Nguyen-Q-Rieu 1981;
\cite{west83b}).
 
\begin{figure*}[pt]
\vspace{8cm}
\includegraphics{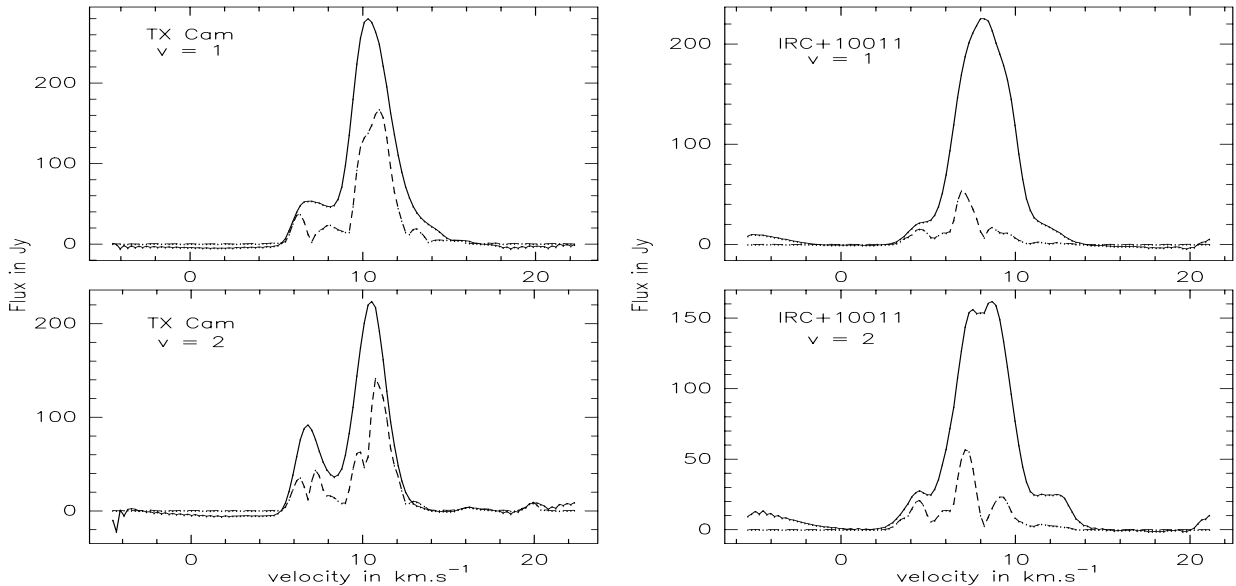}
\caption{Auto correlation (continuous line) and crosscorrelation (dash)
spectrum of TX\,Cam and IRC\,+10011 for the $v$=1 and $v$=2 masers   
                                                  \label{fig_auto_cros} }
\end{figure*}

\begin{figure*}[pt]
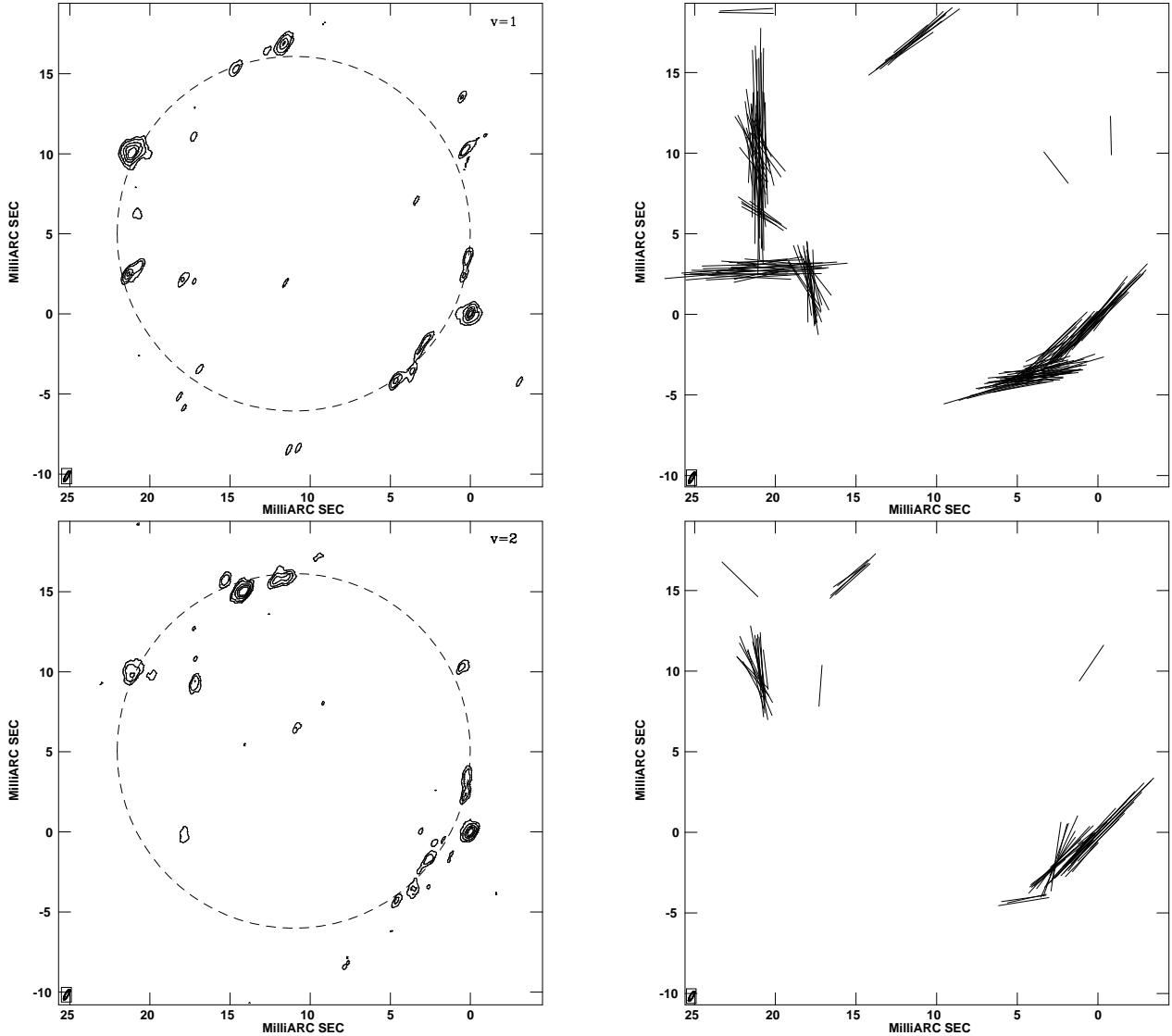

\vspace{15cm}
\includegraphics{h1543.f2a}
\includegraphics{h1543.f2b}
\includegraphics{h1543.f2c}
\includegraphics{h1543.f2d}
\includegraphics{h1543.f2e}
\caption{Integrated intensity (left) and polarization maps (right) of
the $v$=1 (top) and $v$=2 (bottom) $J$=1--0 lines of SiO towards
IRC~+10011.  Contours are multiples by 10\% of the peak flux in each
transition (68~Jy and 67~Jy, respectively). The vectors in the linear
polarization maps indicate the plane of the electric field vector and
their length are scaled as 1 mas $\equiv$ 0.5~Jy~beam$^{-1}$.  The
dashed circles are identical and are drawn to ease the comparison of
both images                                        \label{fig_irc} }
\end{figure*}


\begin{figure}[pt]
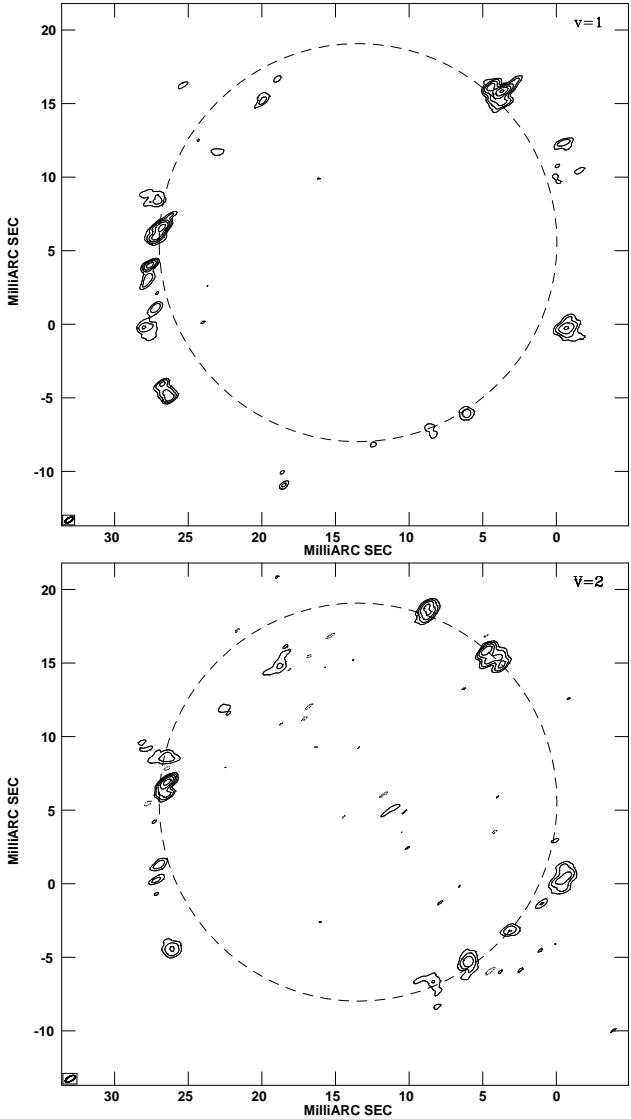

\vspace{15cm}
\includegraphics{h1543.f3a}
\includegraphics{h1543.f3b}
\includegraphics{h1543.f3c}

\caption{Integrated intensity maps of the $v$=1 (top) and $v$=2
(bottom), $J$=1--0 lines of SiO towards TX~Cam.  Contours are multiples
by 10\% of the peak flux in each transition (96~Jy and 72~Jy,
respectively).
The dashed circles are identical and are drawn to ease the comparison
of both images                                  \label{fig_txcam} }
\end{figure}

\begin{figure}[pt]
\vspace{9cm}
\includegraphics{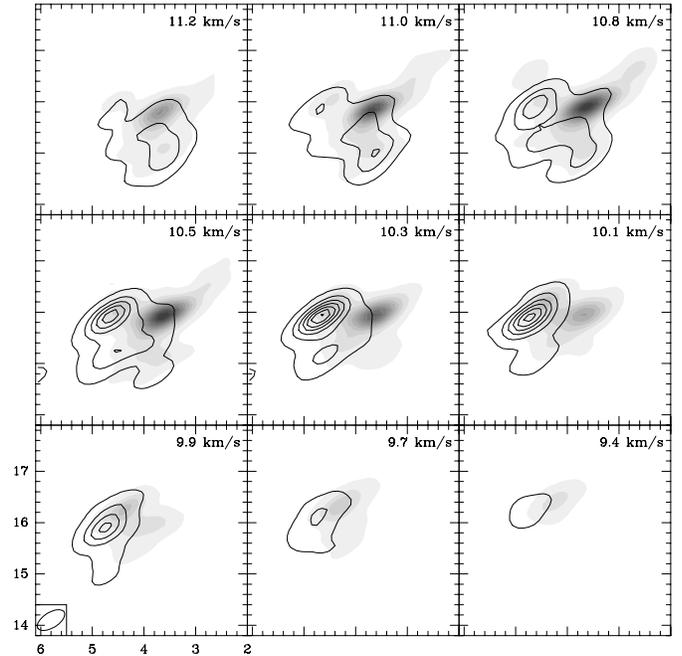}
\caption{Intensity channel map of the maser features at a position
offset of $(5,15)$~mas in TX~Cam, displaying the difference in position
that is systematically observed between the $v$=1 (towards the west,
level in grey) and $v$=2 emission (towards the east, contour levels,
see text). Lowest contour is 5\% of the flux peak (25~Jy/beam)
                                                \label{fig_zoom1} }
\end{figure}


\section{Observations and data analysis}
\label{obs}

We performed observations of the $v$=1 and $v$=2, \mbox{$J$=1--0} lines
of SiO (at a rest frequency of 43,122.080 and 42,820.587 MHz
respectively) with the NRAO Very Long Baseline Array\footnote{The
National Radio Astronomy Observatory (NRAO) is operated by Associated
Universities, Inc., under cooperative agreement with the National
Science Foundation.} in April 1996. The system was setup to record 4
MHz of left-circular and right-circular polarizations in both lines.
The correlation was produced at the VLBA correlator in Socorro (NM,
USA) providing 128 spectral channel cross correlations of the parallel-
and cross-hand polarization bands (thus achieving a spectral resolution
of $\sim 0.22$~km\,s$^{-1}$), out of which all four Stokes parameters
could be retrieved, and hence the full polarization state of SiO
masers.

The calibration procedure used to derive the total intensity maps
followed the standard scheme in the Astronomical Image Processing
System (AIPS) for spectral line experiments.  Maps of $50\times50$ mas
were produced for both $v$=1 and $v$=2 lines separately, with a
restoring beam of $0.6\times0.3$~mas FWHM (with the major axis being at a
position angle PA = $-60^\circ$, measured from North to East) for
TX~Cam and $0.7\times0.2$~mas (PA = $-29^\circ$) for IRC~+10011.

AIPS also provides a set of tools to produce maps of the linearly
polarized spectral line emission (for more details see e.g. the AIPS
Cookbook\footnote{The AIPS Cookbook is available at the URL
http://www.cv.nrao.edu/aips/cook.html}, Cotton 1993, Leppanen et
al. 1995, and Kemball et al. 1995). We observed the quasar 3C~454.3
(of known linear polarization) to estimate the delay correction between
the RCP and LCP bands. The instrumental polarization leakage (or
D-terms) were estimated from observations of the unpolarized quasar
3C~84, consistently with the corrections found independently with AIPS
from the SiO maser sources.

In Figs$.$~\ref{fig_auto_cros}, we show the spectrum of
auto-correlation and cross-correlation. These spectrum tell us that in
the case of TX~Cam less than 50\% of the flux is lost in our map due to
our high spatial resolution, but for IRC~+10011, about 75\% of the flux
is lost.  This suggest that the map of IRC~+10011 is essentially
sensible to the most compact components and that we missed part of flux
due to the presence of extended components.

The determination of the absolute polarization position angle for
IRC~+10011 was performed using single dish observations made at the
same epoch with the radio telescope at Centro Astron\'omico de Yebes
(OAN, Spain).  We identified and measured the polarization angle of a
component spatially isolated within our velocity resolution in our VLBA
maps(Desmurs et al. 1999). We measured the absolute position angle of
the polarization observed at the same velocity in our single dish
data. Comparing both polarization angles, we were able to find the
correction to be applied to our VLBI map in order to be in agreement
with our single dish observations. We estimate that the confidence in
the final position angle is $\pm 10^\circ$.  We checked also that the
correction obtained in this way, applied on the polarization map of
3C~454.3, produced a result consistent with that shown by Kemball et
al. (1996).

Maps of the two transitions were independently produced by solving the
residual fringes-rates on the line sources.  This correction is
determined by selecting a channel containing a simple feature with a
simple spatial structure and a high S/N ratio (strong emission), that
is used as phase reference for all other channels. To do this, we
selected a maser spot appearing in both transitions assuming that these
spots were actually spatially coincident.  This fixed the position of
that maser spots at the map origin in both transitions. This procedure
assumes that this maser spot that appears in both lines at the same
velocity and position respectively to the rest of the emission
distribution actually arises from the same    condensation (see result on
Figs$.$~\ref{fig_irc} and \ref{fig_txcam}); a similar procedure was
followed for example, by Miyoshi et al. (1994).


\section{Results and discussion}
\label{results}

Our high-quality VLBA maps (Figs$.$~\ref{fig_irc} and \ref{fig_txcam})
confirm that both $v$=1 and $v$=2 $J$=1--0 maser transitions arise from
ring-like structures, at a distance from the center $\sim 11$~mas for
IRC~+10011, and $\sim 14$~mas for TX~Cam. These radii are equivalent to
about $8\cdot 10^{13}$~cm (assuming a distance of $\sim 500$~pc for
IRC~+10011, and $\sim$ 350~pc for TX~Cam). In IRC~+10011, a well
defined pattern of tangential linear polarization is found. Linear
polarization degrees as high as 22\% are measured, though some spots
are not linearly polarized.  A tangential polarization pattern has been
also detected in TX~Cam, consistent with the work published by Kemball
et al. (1995), and with similar properties to those of we find in
IRC~+10011.

The above observational results are expected from theoretical models
for SiO radiative pumping. In fact, both the flux distribution in a
thin ring and the tangential polarization are properties essentially
related to the radiative pumping of SiO masers.
Excitation of SiO vibrational states by absorption of stellar photons
is much more probable than collisional excitation, as one can easily
demonstrate by taking into account the expected physical conditions in
the emitting regions. For instance, an SiO shell located at 3 stellar
radii, with a kinetic temperature of 1500~K and a density of 10$^9$
particules cm$^{-3}$, presents an absorption rate of stellar photons in the
$v$=0$\rightarrow$1 transition that is about 200 times larger than the
corresponding collisional excitation rate. However, radiative
excitation is efficient to produce population inversion in $v > 0$
rotational transitions only when the SiO region fulfills some
conditions, particularly when radiation comes from a central source at
some distance from the SiO shell (Bujarrabal \& Nguyen-Q-Rieu 1981,
Bujarrabal 1994a).  Under such conditions, the high probability of the
photon absorption implies that the pumping must be radiative. Due to
the geometrical conditions for the radiative pumping, the radiative
maser excitation is more efficient close to the inner boundary of the
SiO shell. In terms of flux, this effect is strengthened by the
non-linear maser amplification (Bujarrabal 1994b, Western \& Watson
1983a). This theoretically indicates that ``tangential'' (i.e$.$
perpendicular to the radial direction) amplification is dominant, and
leads to a radial flux distributions in a thin ring, strikingly similar
to that observed in these stars.

Tangential polarization is also expected from radiative pumping models,
as can be understood from a semiclassical point of view.  Radiation
incident on a gas shell from a central source must show electric and
magnetic fields perpendicular to the radial direction. Linear molecules
are preferentially excited by the absorption of such photons when they
rotate in the tangential plane, and will preferentially rotate in it
after the excitation. This is expected to be the case when SiO is
vibrationally excited from the $v$ to the $v$+1 state by absorption of
stellar 8$\mu$ photons. In other words, for a given $J$ level in $v$+1,
the most populated magnetic sublevels will be those with largest $|M|$
values (when the quantization of the magnetic quantum number is taken
along the radial direction). When the molecular emission is
essentially tangential (as for SiO masers), the tangential plane, in
which molecules rotate, contains the line of sight for intense spikes:
one then expects to detect ``tangential polarization'', in agreement
with observations. Note that this mechanism does not require any
magnetic field or exotic maser saturation effect.

This intuitive explanation of the linear polarization observed in SiO
masers was first proposed, to our knowledge, by Bujarrabal \&
Nguyen-Q-Rieu (1981). Soon later, but well before the first VLBA
polarization maps, Western and Watson (1983a,b) developed more in
detail the theory of maser polarization under these conditions,
confirming the expected presence of strong tangential polarization in
the case of SiO masers from evolved stars. We can also see a more
recent discussion in Elitzur (1996).

\begin{figure}[pt]
\vspace{12cm}
\includegraphics{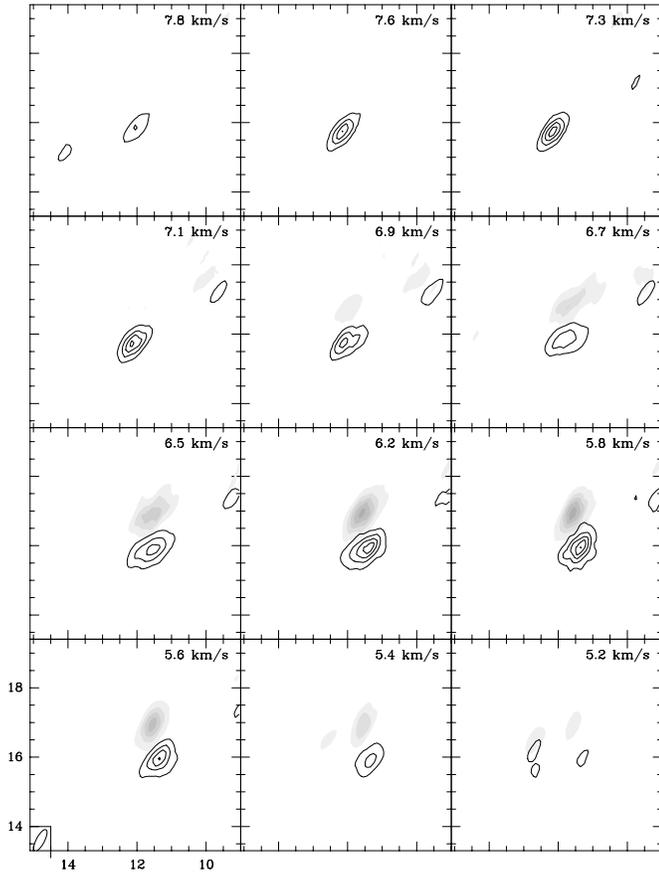}
\caption{Same as Fig.~\ref{fig_zoom1}, but for the maser features at a
position offset of $(+12, +16)$~mas in IRC~+10011. Lowest contour is 5\%
of the flux peak (11.5~Jy/beam) \label{fig_zoom2} }
\end{figure}

Our maps also show that the $v$=1 and $v$=2 spots are systematically
not coincident. 
Positioning of the $v$=1 features relative to the $v$=2 ones has been
performed as described in Sect.~\ref{obs}.
In TX~Cam we identify, from the integrated flux maps in
Fig.~\ref{fig_txcam}, about 17 features in both $v$=1 and $v$=2
transitions, out of which only one appears to be spatially coincident,
within the resolution of our observation. In the case of IRC~+10011, we
find about 13 features, 4 of which could be coincident (see
Fig.~\ref{fig_irc}). In most cases in which spots of these lines are
nearby, there is a systematic shift between them, of about 1--2~mas
($\sim$ 10$^{13}$~cm), being the $v$=2 ones placed in a layer slightly
closer to the star (Figs.~\ref{fig_zoom1} and \ref{fig_zoom2}).

The way how the maps in the two lines are aligned (see Sect. 2) implies
that at least one maser spot appears in both lines at the same spatial
position respectively to the rest of the emission distribution.
Nevertheless, we can easily check that there is no possibility of
selecting any spot that permits the alignment of a majority of the
other spots in both lines.

Another geometrical argument is that the diameters of both emission
rings differ. After suming all the velocity channels into one velocity
integrated map (Fig.~\ref{fig_irc} and \ref{fig_txcam}), we determined
the position of the maser spots showing a S/N ratio at least equal to 5
$\sigma$ (for TX~Cam 1$\sigma$ is $\sim$0.86 Jy and for IRC~+10011
1$\sigma$ is $\sim$0.7 Jy).
We fited a ring shell to our data giving a relative weight to each maser
spot proportional to its intensity.  The parameters (position center and
radius of the emission ring) for which the coefficient of standard
deviation is minimum is kept as the best possible fitting.
As a result, for TX~Cam the radii found for the $v$=1 and
$v$=2 maser transition are respectively 14.5$\pm$0.2 mas and
13.2$\pm$0.2 mas. This represents a difference between both diameters
of the emission rings of 2.6 $\pm$0.5 mas, to be compared with our
beam resolution of 0.6$\times$0.3 mas.
For IRC~+10011, the radii found are 11.7$\pm$0.2 mas for the $v$=1 and
11.1$\pm$0.2 mas for the $v$=2, which gives a diameter difference of
1.2$\pm$0.5 mas, for a resolution beam of about 0.7$\times$0.2 mas.
These differences in the maser ring sizes are significant enough in
comparison with our beam resolution, which leads us to conclude that
the two lines come from shells with significantly different diameters.

The conclusion that the $v$=1 and $v$=2 masers arise from the spatially
coincident spots obtained previously by Miyoshi et al. (1994) is then
not confirmed by our higher-resolution data. This disagreement could be
explained by the fact that the measured shifts are significantly
smaller than the beam in their data (7~mas).  The claim by these
authors ruling out radiative pumping, on the basis that radiative
models tend to require different conditions for the pumping of
transitions in these two vibrational states (see Sect.~\ref{intro}),
was clearly premature. In fact, the systematic separation between the
$v$=1 and $v$=2 spots in our data must be considered as an additional
support for the radiative pumping.

Preliminary calculations for radiatively pumped models (see
\cite{buja94a} and \cite{herp98}) indicate that, when both $v$=1 and 2
masers appear in the same clump, the $v$=2 lines tend to select its
innermost parts, in agreement with the data. In any case, published
calculations have not explicitly addressed this important point yet,
and new accurate calculations are required.

Our observational results could also be explained under certain
condition by collisional pumping. The tangential polarization can be
produced in the presence of strong magnetic fields within certain
directions, independently of the pumping process. But this mechanism
cannot explain the tangential polarization if it is (as it seems) a
systematic property in evolved stars.  For instance, Kemball \& Diamond
(1997) argue that a pattern like that detected in TX Cam by them would
require a poloidal magnetic field with a non-arbitrary inclination of
the pole with respect to the plane of the sky, $\sim$
40$^\circ$--60$^\circ$, other angles even leading to radial
polarization.  Shell-like spatial distributions of the maser flux can
also appear when the pumping is collisional, if it acts in certain
pulsating atmospheres (\cite{hump96}, \cite{doel95}). As we have
mentioned, the collisional pumping models tend to predict spatial
coincidence between the $v$=1 and $v$=2 masers, though in the above
mentioned papers there are also examples in which the $v$=2 $J$=1--0
masers are placed closer to the star.

But the main conclusion from the discussion presented in this paper is
that the bulk of the existing empirical results on SiO maser emission
from evolved stars strongly support radiative pumping. The spatial
shell-like distribution of the flux, the tangential polarization
pattern of $v$=1 and 2 $J$=1--0 SiO masers, and the shift in their
spatial distributions (properties very often found in the existing
observations) are intrinsically related with the radiative pumping
schemes. Indeed, radiative models predicted, well before the
observations were made, that such properties should be systematically
observed.

\begin{acknowledgements}
We thank an anonymous referee for helpful comments that improved the
manuscript.
This work has been partially supported by the Spanish DGES Project
PB96-0104.  JFD acknowledges support for his research by the European
Union under contract ERBFMGECT950012. 
\end{acknowledgements}


\end{document}